\documentclass[aps,prl,twocolumn,showpacs]{revtex4}

\newcommand{\Jeff}{J_\mathrm{eff}}

\usepackage{graphicx,color}

\begin{document}

\title{Generation of uniform synthetic magnetic fields
by split driving of an optical lattice}

\author{C.E.~Creffield and F.~Sols}
\affiliation{Departamento de F\'isica de Materiales, Universidad
Complutense de Madrid, E-28040, Madrid, Spain}

\date{\today}


\begin{abstract}
We describe a method to generate a synthetic gauge
potential for ultracold atoms held in an optical lattice.
Our approach uses a time-periodic driving potential 
based on quickly alternating two Hamiltonians to
engineer the appropriate Aharonov-Bohm phases, and permits
the simulation of a uniform tunable magnetic field.
We explicitly demonstrate that
our split-driving scheme reproduces the behavior of a charged quantum
particle in a magnetic field over the complete range of
field strengths, and obtain the Hofstadter butterfly
band-structure for the Floquet quasienergies.
\end{abstract}

\maketitle

{\em Introduction -- }
Systems of ultracold atoms held in optical lattice
potentials have an exceptionally high degree of controllability and
excellent coherence properties. As such they
have proven to be excellent systems for simulating \cite{bloch_sim}
Hamiltonians arising in diverse areas of physics,
such as graphene \cite{wuns08,graphene}, Majorana fermions \cite{majorana},
and models of high-temperature superconductivity \cite{hubbard}.
A topic of intense current activity
is how to reproduce the effects of
gauge fields in these systems. This would not only
extend the use of cold-atom simulators to new domains, but is also of
considerable interest to applications such as quantum
computation. A particularly important
example is the $U(1)$ gauge of electromagnetism. The ability
to simulate magnetic fields would give exciting new
ways to explore quantum Hall physics and related effects such as
topological insulators and anyon physics, 
together with the realization of
phenomena such as the Hofstadter
butterfly \cite{hofstadter},
a fractal energy spectrum
seen in lattice systems exposed to high magnetic fluxes.

Many efforts to simulate an applied magnetic field
with cold atoms have used laser-driven transitions
between internal atomic states to generate phases
\cite{raman,jaksch,lewenstein} which
mimic the Aharonov-Bohm phases that would be experienced by
charged particles moving in a uniform magnetic field.
An attractive alternative to these schemes
is to use inertial forces, which do not require
a specific internal state structure, and so are applicable
to a wider range of atomic species.
First efforts of this kind \cite{coriolis} used rotation to generate a
Coriolis force, which has an analogous form to the Lorentz force
of electromagnetism. Only weak fields, however, were accessible
by this method. 

An alternative inertial approach
is to periodically accelerate (or ``shake'')
the lattice to produce the effect known as coherent destruction
of tunneling, a quantum coherent effect in which the driving
renormalizes the tunneling amplitude \cite{CDT}. This effect has been
directly observed in the expansion dynamics of atomic clouds
\cite{arimondo_J0,arimondo_Jn,cec_morsch}. More recently, 
it has also been noted that the
driving can be used to render the tunneling complex
\cite{cec_J0,cec_phase,longhi}, giving the prospect of generating
synthetic magnetic fields.
This possibility has been explored experimentally first
in one-dimensional lattices \cite{eckardt1}, and later in a
triangular lattice where
this effect was used to create a staggered field \cite{eckardt2}.

Generating a {\em uniform} field on a {\em square} lattice, however, 
is not straightforward, and an initial proposal \cite{kolovsky}
was later found to contain a problem that limited it
to only inhomogeneous fields \cite{comment}.
Very recently experimental progress has been made \cite{ketterle,bloch}
by introducing additional lattice
potentials and a strong magnetic field gradient.
In this work we address the problem in a different way by borrowing the
well-known split-operator technique from quantum simulation
to develop an alternative method that we term ``split driving''.
This consists of
dividing the time-dependent Hamiltonian into two parts,
namely tunneling in the $x$ and $y$ directions, and
applying the parts sequentially. We show that this avoids the 
problems encountered previously \cite{kolovsky}, and indeed
makes it possible to
simulate a practically uniform synthetic field of arbitrary strength 
using {\em only} the manipulation of the
time-dependent driving potential, without requiring
the complication of external magnetic fields. Furthermore, as
our scheme can work with small driving amplitudes, heating
effects can be minimal \cite{ketterle}, enhancing the chance to
observe the Hofstadter butterfly \cite{troyer}.

\begin{figure}
\begin{center}
\includegraphics[width=0.45\textwidth,clip=true]{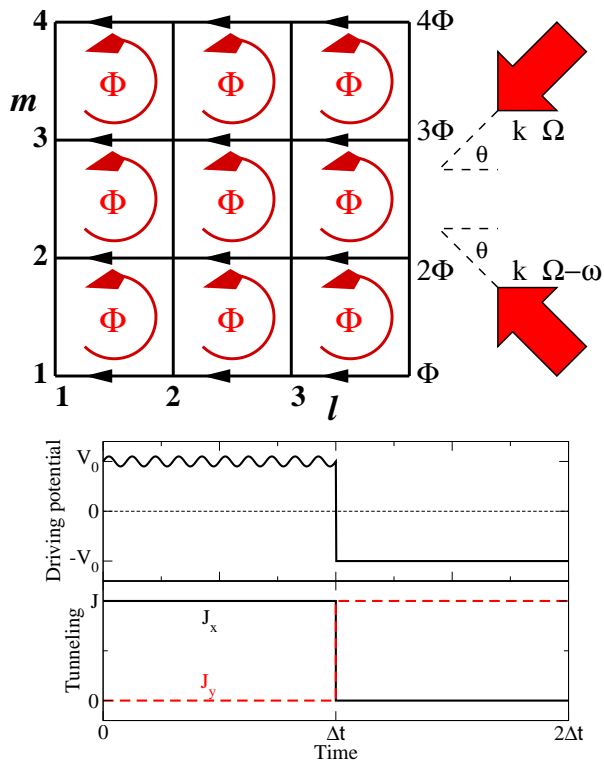}
\includegraphics[width=0.40\textwidth,clip=true]{fig1b}
\end{center}
\caption{(a) Schematic representation of the method.
The optical lattice is formed as a two-dimensional
standing wave between incoming laser beams, and is periodically accelerated
in the $x$ direction, producing an inertial force
in the lattice rest-frame.
Two additional running-wave beams, with a frequency difference
$\omega$ and crossing at an angle of $2 \theta$, add a spatially-dependent 
phase to the driving. In the Landau gauge the tunneling phases $\chi$
appear only on the horizontal hopping processes (marked with arrows).
A linear variation of the phase with $y$,
$\chi(m) = m \Phi$, would produce
a uniform flux $\Phi$ threading each plaquette.
(b) The split-driving scheme. In the first half-period,
an acceleration consisting of a constant and
oscillating component is applied in the $x$-direction (upper figure)
to induce the Aharonov-Bohm tunneling phases,
while the tunneling in the $y$ direction ($J_y$) is suppressed (lower figure).
In the second half-period the $x$-coordinate is uniformly accelerated
back to its original position, suppressing the $x$-tunneling ($J_x$) via the
Wannier-Stark effect,
while tunneling in the $y$ direction is restored. This pattern of driving
is periodically repeated.}
\label{scheme}
\end{figure}

{\em Method -- }
To illustrate the main ideas we schematically show a possible
arrangement in Fig.\ref{scheme}a.
We consider a two-dimensional optical lattice, formed by
the superposition of two orthogonal standing waves.
When the optical lattice is sufficiently deep,
a system of cold atoms can be described well by a tight-binding
(hopping) Hamiltonian
\begin{equation}
H(t) = -J \sum_{ i} \left(
a_i^\dagger a_{i + {\hat x}} + a_i^\dagger a_{i + {\hat y}} + 
\mathrm{H.c.}  \right) +
H_I(t)
\label{hamiltonian}
\end{equation}
where $J$ is the hopping between nearest-neighbors
and $a_i \ / a_i^\dagger$ are the
standard particle annihilation/creation operators for lattice site $i$.
Acceleration of the lattice in the $x$-direction
can be viewed in the rest frame of the lattice as 
arising from an inertial force,
described as a scalar potential depending linearly on $x$,
$H_I(t) = V(t) \sum_j x_j n_j$ where
$x_j$ is the $x$-coordinate of lattice-site $j$ and
$n_j$ is the number operator. We take the
specific form for the driving 
\begin{equation}
V(t) = V_0 + K \cos \left(\omega t + \alpha \right) \ ,
\label{potential}
\end{equation}
consisting of a constant and an oscillation, parameterized by
$V_0$ and $K$ respectively.
The frequency of the oscillation is $\omega$, and
$\alpha$ is the driving phase-shift which will play
an important role. The driving starts at $t=0$. We 
assume that the resulting discontinuity at $V(t = 0)$ 
is experimentally feasible, due to the relatively
slow dynamics of cold atoms.

We begin by considering the simplest case of constant $\alpha$, and throughout
this work we shall work with {\em resonant driving},
$V_0 = n \omega$ where $n$ is an integer, 
and we take $\hbar = 1$.
In the limit of high frequencies, $\omega \gg J$,
the system can
be described by an effective static hopping Hamiltonian, with
an effective $x$-hopping given by \cite{cec_phase}
\begin{equation}
\Jeff = J \ e^{i \left( K / \omega \right) \sin \alpha} \ 
e^{i n \left( \alpha + \pi \right)} \
{\cal J}_{n} \left( K / \omega \right) ,
\label{effective_J}
\end{equation}
for right-to-left hopping, and $\Jeff^\ast$ for left-to-right movement.
Note that for $\alpha=0$ this reduces to the well-known Bessel
function renormalization \cite{holthaus_Jn},
$\Jeff = (-1)^n J {\cal J}_{n} \left(K / \omega \right)$.
From Eq. (\ref{effective_J}) we can clearly see that the driving
in general produces a tunneling-phase \cite{cec_phase},
$\chi = \left( K / \omega  \right) \sin \alpha + 
n ( \alpha + \pi)$.
For one-dimensional systems
this has been proposed as a means to induce transport 
\cite{cec_phase,cec_J0},
and the effects of $\chi$ have been experimentally observed in 
experiments on driven
``super-Bloch oscillations'' \cite{innsbruck,tania_phase}.

The magnetic flux passing through a plaquette is evaluated
by summing the complex phases acquired on each tunneling process as the
plaquette is traversed anti-clockwise. 
Clearly when $\chi$ is constant the flux is zero \cite{footnote_phase}, 
and so to produce a synthetic flux $\chi$ must vary in space.
The specific
case of the Landau gauge $\chi (m) = m \Phi$,
where $m$ indexes the $y$-coordinate, is shown in
Fig.\ref{scheme}a, which results in every
plaquette being threaded by a net flux $\Phi$.
To mimic this gauge we therefore make the
phase-shift similarly spatially-dependent,
$\alpha \rightarrow \alpha(m) = m \phi$.
From Eq. (\ref{effective_J}) this gives an effective
flux per plaquette of
\begin{equation}
\Phi(m) = n \phi + \left( K / \omega \right) \left[
\sin \left( \left( m+1 \right) \phi \right) \ - \
\sin \left(  m \phi \right) \right] \ .
\label{net_flux}
\end{equation}
The first term of this expression indeed reproduces the
Landau gauge of a uniform synthetic flux, while the second 
corresponds to a flux that
varies with $y$. Its value, however, is bounded,
$| \Phi_{\mathrm{var}} | \leq 2 
\left| \left(K/\omega \right)\sin \left( \phi/2 \right) \right|$, 
and so can be controlled
by setting $K/\omega$ sufficiently small. 
Besides generating an essentially uniform $\Phi$,
making $K/\omega$ small has the additional advantage
of limiting lattice heating effects,
although as this reduces the amplitude of $\Jeff$ this also has
the effect of slowing the system's dynamics.

In an experimental realization, we thus need to introduce two vital ingredients;
a time-dependent driving potential to produce photon-assisted tunneling
in the $x$ direction, and a $y$-dependence of the phase, $\alpha(m)$,  that
is imprinted on this tunneling. This can be done in a number of different
ways, and in the particular scheme shown in Fig. \ref{scheme}a it 
is produced by a pair
of far-detuned running wave beams
with wavevector $k$ and a frequency difference of $\omega$ between them.
This produces a  time-dependent optical potential 
$V_{\mathrm{RW}}(x) \cos^2 \left( \omega t/2 + k y \sin \theta \right)$,
where $2 \theta$ is the angle between the beams 
and $V_{\mathrm{RW}}(x)$ is the envelope function of the light intensity.
If this varies weakly with $x$ then to first order we can write \cite{kolovsky}
\begin{equation}
V_{\mathrm{RW}}(x) \simeq V_{\mathrm{RW}}(0) + 
x \frac{\partial V_{\mathrm{RW}}}{\partial x} \ ,
\label{taylor}
\end{equation}
where the first term is an unimportant constant, and the second produces
the oscillatory component of the driving potential.
In combination with a uniform acceleration of the optical lattice, this
gives a potential which is a generalization of Eq.\ref{potential} 
in which the phase, $\alpha$, varies with position
\begin{equation}
V(m,t) = V_0 + K \cos \left( \omega t + \alpha(m) \right) \ ,
\label{Vspace}
\end{equation}
where $\alpha(m)  = 2 k d_L m \sin \theta$
and $d_L$ is the optical lattice spacing
(henceforth we take $d_L = 1$), 
and $K =V'_{\mathrm{RW}} / 2$.
In practice the gradient $V'_{\mathrm{RW}}$ does not have to be strictly 
constant over the entire cloud, as long as its variation is 
sufficiently weak that Eq. \ref{taylor} is valid.
Interference between the optical lattice
and the running-wave beams can be avoided by using
acoustic-optic modulators to produce
frequency offsets, as described in Ref.\cite{ketterle}. 
Unfortunately, when $\alpha(m)$ varies in this way,
the potential (\ref{Vspace})
has the undesired effect of also driving tunneling in the $y$-direction
since the potential difference between
a site and its neighbors in the $y$-direction,
$V(m \pm 1,t)-V(m,t)$, is in general a time-dependent
quantity, oscillating with frequency $\omega$ \cite{comment}.

To eliminate this unwanted renormalization of the
$y$-tunneling, we apply the
split-operator technique \cite{soren} familiar from numerical studies
of quantum systems, and divide the time-evolution operator
over a short time-interval into two parts,
as shown in Fig. \ref{scheme}b.
In the first interval the lattice is driven by the potential
(\ref{Vspace}), while the $y$-hopping is suppressed
(for example, by increasing the depth of the optical lattice in the
$y$-direction). In this interval $H(t)$ can thus be replaced
by an effective static Hamiltonian, $H_x^{\mathrm{eff}}$,
which contains only $x$-hopping terms in which the tunneling has been
renormalized.
In the second interval we restore the $y$-hopping, and instead
suppress the $x$-hopping, so that the system evolves under
$H_y$, a time-independent Hamiltonian containing only the $y$-hopping operators.
A convenient way to do this
is to flip the sign of the acceleration of the lattice
($V_0 \rightarrow -V_0$ while setting $K=0$) so that the
intersite tunneling in the $x$-direction is destroyed by Wannier-Stark
localization. This has the additional practical benefit of keeping the average
displacement of the lattice zero \cite{footnote}, otherwise the constant
acceleration would quickly move the lattice out of the
experimental area.

This division amounts to a Suzuki-Trotter decomposition of $H(t)$
\begin{equation}
e^{-i \left( H_x^{\mathrm{eff}} + H_y \right) \Delta t}
\simeq
e^{-i H_x^{\mathrm{eff}} \Delta t}
e^{-i H_y \Delta t} \ .
\label{trotter}
\end{equation}
As $H_x^{\mathrm{eff}}$ and $H_y$ do not commute, the leading error in this
decomposition is given by the Baker-Campbell-Hausdorff formula
$\frac{\Delta t^2}{2} \left[H_x^{\mathrm{eff}}, H_y \right] \sim
J \Jeff \Delta t^2$.
More complicated decompositions can be used in which the error
term decays more rapidly \cite{suzuki}, but for simplicity we limit
ourselves to the most primitive form.
Henceforth we set the tunneling in the $y$ direction $J_y = \Jeff$,
so that the error terms arising in the two time intervals,
$(J \Delta t)$ and $(\Jeff \Delta t)$ are equal. If $\Jeff \neq J$
the same effect can be produced by taking the two time intervals
to be of different lengths.
For good accuracy we must take $\Delta t$ to be as small
as possible, but to obtain the effective renormalization of the tunneling
(\ref{effective_J}) $\Delta t$ must be larger than the driving
period $T=2 \pi/\omega$. Numerically we have found the minimum
period to be $\Delta t = 3T$; below this value the renormalization
effect is abruptly lost. 
All the results we show below are
obtained for a time-interval of $\Delta t = 8 T$.

\begin{figure}
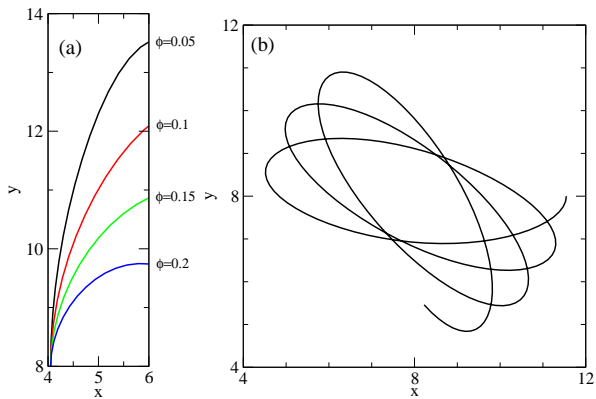

\begin{center}
\includegraphics[width=0.14\textwidth,clip=true]{fig2a}
\includegraphics[width=0.29\textwidth,clip=true]{fig2b}
\end{center}
\caption{Weak field results.
(a) The system was initialized as a Gaussian wavepacket of width $\sigma^2 = 5$, and kicked in the
$+y$ direction at $t=0$ by applying a phase imprinting $\exp \left(i y \right)$.
The center of mass of the wavepacket describes a circular orbit, the
radius of which is inversely proportional to the effective flux, 
in analogy to the cyclotron orbit of a charged particle in a
uniform magnetic field.
(b) Here the system is initialized in the ground state of
a parabolic trap potential $V = \kappa r^2/2$, with curvature $\kappa = 0.1 J$, 
which is then shifted 4 lattice spacings to
the left. The center of mass traces out a rosette pattern, precisely
analogous to the path of a Foucault pendulum subjected to the
Coriolis force. Physical parameters: driving frequency $\omega=50J$,
driving amplitude $K = 0.05 \omega$, and $n = 1$.}
\label{weak}
\end{figure}

{\em Results -- }
At low values of $\Phi$, we can understand the behavior
of the system semiclassically. We consider a $15 \times 15$ lattice
with open boundary conditions, initialize the system
as a narrow Gaussian wavepacket, and propagate it in
time using the time-dependent Hamiltonian
(\ref{hamiltonian}) and the split-driving protocol
with $n=1$ and  $K/ \omega$ small enough for the variation
in $\Phi$ to be negligible, implying $\Phi = \phi$.
In Fig. \ref{weak}a
we show the effect of giving the initial state
a kick in the $+y$ direction at $t=0$ for various
values of $\phi$. We can clearly see
that in each case the center of mass follows a circular trajectory,
analogous to the Larmor orbit of a classical particle under
the Lorentz force. As the synthetic magnetic flux is increased, the radius
of the orbit decreases proportionately, as expected.
Since the wavepacket contains a number of different quasimomenta,
it spreads during the time-evolution, however, and eventually contacts
the edge of the lattice, distorting the circular motion of the
center of mass.

In Fig. \ref{weak}b we show the time evolution of a
Gaussian wavepacket in the presence of a parabolic 
trap potential \cite{kol_new},
for a small $\Phi = 0.02$. At $t=0$ the
trap potential is abruptly shifted 4 lattice spacings to the left, thereby
exciting the wavepacket into motion. In the absence of the magnetic
flux, the wavepacket would simply slosh from one side of the trap to the other
\cite{slosh}. However, with the flux present the wavepacket
experiences a Lorentz force perpendicular to its direction of motion, causing its
center of mass to trace out the characteristic rosette pattern
seen in Fig. \ref{weak}b. This is precisely analogous
to the path traced by a Foucault pendulum under the influence of
the Coriolis force; in a reversal of the procedure used in
Ref. \onlinecite{coriolis} we can thus
use the synthetic gauge field to mimic the effects of rotation.
Due to the confining harmonic potential, the wave packet stays away from
the lattice edges, thus displaying close to ideal behavior even for long times.

\begin{figure}
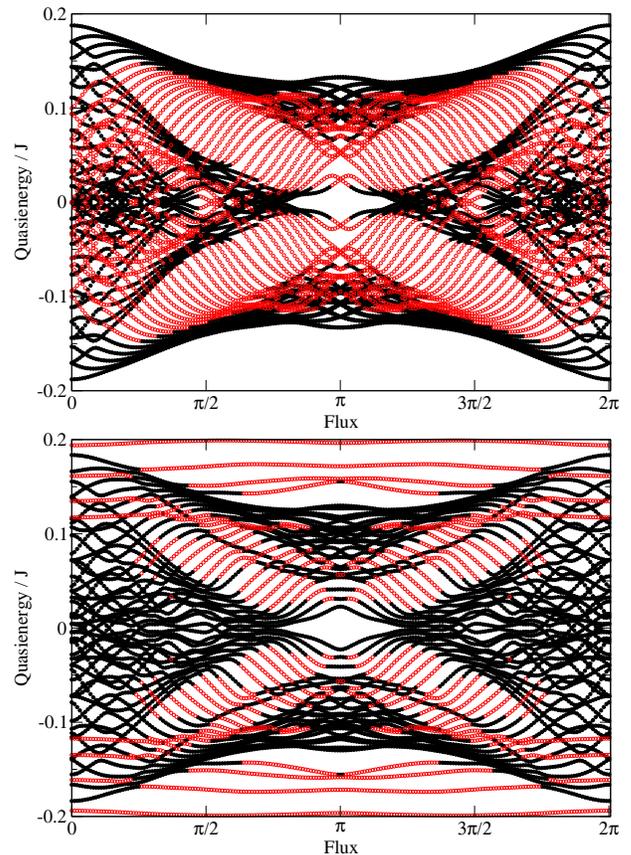

\begin{center}
\includegraphics[width=0.45\textwidth,clip=true]{fig3a}
\includegraphics[width=0.45\textwidth,clip=true]{fig3b}
\end{center}
\caption{Quasienergy spectrum for high and low frequencies;
black (solid) circles denote the bulk states, red (hollow) circles
are states with more than 50\% of their density localized on the edge.
Driving parameters are $n=1$ and $K=0.1 \omega$.
(a) $\omega = 1000J$, for this high frequency, the spectrum is
indistinguishable from the ``butterfly'' spectrum of the
true Hofstader-Harper Hamiltonian.
(b) $\omega = 10J$, some differences appear between the quasienergy
spectrum and the Hofstadter butterfly; fewer edge states are visible,
and a subset of edge states shows only a weak dependence on
the magnetic flux.}
\label{butterfly}
\end{figure}

For larger values of flux, when the magnetic length is
comparable or less than the lattice spacing, the system's
behavior shows a complicated quantum interference pattern
and it is no longer possible to use semiclassical intuition to
understand it.
As the Hamiltonian of the system is time-periodic, its dynamics is 
governed by its 
quasienergies $\epsilon_j$ \cite{hanggi}. These are a generalization of the 
energy eigenvalues familiar from static systems, 
related to the eigenvalues of the time-evolution operator for one period
$U(t+T,t)$ via $\lambda_j = \exp \left[-i T \epsilon_j \right]$. 
In Fig. \ref{butterfly}a we show the quasienergy
spectrum of the driven Hamiltonian for an
$8 \times 8$ lattice as $\Phi$ is varied
from 0 to $2 \pi$. In the upper figure the system is
driven at a very high frequency $\omega = 1000J$. The quasienergies
clearly have the form of the Hofstadter butterfly,
and indeed this spectrum is indistinguishable from the
energy spectrum of the true (static) Harper-Hofstadter Hamiltonian
\cite{hofstadter}. We can distinguish
two kinds of states. 
If more than $50 \%$ of the wavefunction density
is located on sites on the perimeter of the lattice
we term the state an ``edge state'', and otherwise
it is termed a ``bulk state''.
The bulk states contain
a self-similar series of gaps, which become fractal
in the limit of large system size. The edge states, which would be absent
in an infinite system or a system with periodic boundary
conditions, lie within these gaps, and are chiral transporting states.
It can be clearly seen that for a given value of flux the edge
states occur in pairs with opposite slope, corresponding
to propagation in either the clockwise or anti-clockwise sense
around the boundary of the lattice.

In Fig. \ref{butterfly}b we take the lower, and more experimentally
realistic, value of $\omega = 10J$
to demonstrate that our simulation procedure is robust.
Reducing the value of $\omega$ affects the results in two
ways; firstly the perturbation theory yielding
Eq. (\ref{effective_J}) becomes increasingly less accurate, and
secondly the error term arising from the Suzuki-Trotter
decomposition (\ref{trotter}) becomes more significant.
These two sources of inaccuracy are responsible for the differences
which Fig. \ref{butterfly}b shows with respect to the ideal behavior of Fig. \ref{butterfly}a.
The butterfly structure is still clearly evident
in the results, although 
some differences with the Harper-Hofstadter spectrum
now appear. In particular we can observe that
a subset of the edge states
show an almost flat dependence on the flux,
indicating that they have zero group velocity, and so
are no longer transporting.
Although these differences become more significant
as $\omega$ is reduced further, we have checked that the broad
structure of the Hofstadter butterfly is still reproduced even for
driving frequencies as low as $\omega=2J$.

{\em Conclusions -- }
In summary, we have described a method of using a periodic driving
potential to produce a synthetic gauge field of arbitrary strength.
To achieve this we have developed a {\it split-driving} approach,
in which the desired time-evolution operator is constructed from
a sequence of unitary operations that
separate the $x$ and $y$ degrees of freedom.
Our method is robust and simple, avoids the need for any
additional fields and, unlike schemes based on hyperfine
transitions, it does not require a specific internal structure for
the atoms.
It is also well within the reach of current techniques. For example, 
a typical optical lattice with $d_L \simeq 500$nm and
bare tunneling $J/h \simeq 100$Hz can be driven
at $\omega /2 \pi\sim$1 kHz 
\cite{arimondo_Jn,ketterle,bloch}, giving $\Delta t \sim$10 ms.

This method opens the prospect of using the
single-site addressability and fine experimental control of cold
atom systems to study the quantum Hall effect and topological
insulator systems in a novel way. 
A particularly appealing application
is to ladder systems \cite{belen}, which provide
a convenient bridge between one-dimensional topological insulators
and the full two-dimensional case.
Exciting future developments
would be the generalization of these results to non-Abelian
gauge theories, and to the rapidly
developing field of Floquet topological insulators, in which
the Floquet quasienergy spectrum itself has a
non-trivial topology \cite{demler}.

\bigskip

The authors thank Wolfgang Ketterle, Maciej Lewenstein,
Juliette Simonet,
and Monika Aidelsburger for stimulating discussions. This
research was supported by the Spanish MINECO through
Grant No. FIS-2010-21372.


\begin{thebibliography}{99}

\bibitem{bloch_sim}
{I.~Bloch, J.~Dalibard, and S.~Nascimb\`ene,
Nat. Phys. {\bf 8}, 267 (2012).}

\bibitem{wuns08}
{B.~Wunsch, F.~Guinea, and F.~Sols, New J. Phys
 {\bf 10}, 103027 (2008).}

\bibitem{graphene}
{L.~Tarruell, D.~Greif, T.~Uehlinger, G.~Jotzu, and T.~Esslinger,
Nature {\bf 483}, 302 (2012).}

\bibitem{majorana}
{L.~Jiang, T.~Kitagawa, J.~Alicea, A.R.~Akhmerov, D.~Pekker, G.~Refael,
J.I~Cirac, E.~Demler, M.D.~Lukin, and P.~Zoller,
Phys. Rev. Lett. {\bf 106}, 220402 (2011).}

\bibitem{hubbard}
{D.~Jaksch and P.~Zoller, P. Ann. Phys. {\bf 315}, 52 (2005).}

\bibitem{hofstadter}
{D.R.~Hofstadter, Phys. Rev. B {\bf 14}, 2239 (1976).}

\bibitem{raman}
{Y.~Lin, R.L.~Compton, K.~Jim\'enez-Garc\'ia, J.V.~Porto, and I.B.~Spielman,
Nature (London) {\bf 462}, 628 (2009);
J.~Dalibard, F.~Gerbier, G.~Juzeli{\=u}nas, and P.~\"Ohberg,
Rev. Mod. Phys. {\bf 83}, 1523 (2011).}

\bibitem{jaksch}
{D.~Jaksch and P.~Zoller, New J. Phys. {\bf 5}, 56 (2003);
E.J.~Mueller, Phys. Rev. A {\bf 70}, 041603 (2004);
F.~Gerbier and J.~Dalibard, New J. Phys. {\bf 12}, 033007 (2010).}

\bibitem{lewenstein}
{A.~Celi, P.~Massignan, J.~Ruseckas, N.~Goldman, I.B.~Spielman, G.~Juzeli{\=u}nas,
and M.~Lewenstein, Phys. Rev. Lett. {\bf 112}, 043001 (2014).}

\bibitem{coriolis}
{K.W.~Madison, F.~Chevy, W.~Wohlleben, and J.~Dalibard,
Phys. Rev. Lett. {\bf 84}, 806 (2000);
J.~Abo-Shaeer, C.~Raman, J.~Vogels, and W.~Ketterle,
Science {\bf 292}, 476 (2001).}

\bibitem{CDT}
{F.~Grossmann, T.~Dittrich, P.~Jung, and P.~H\"anggi, Phys. Rev. Lett.
{\bf 67}, 516 (1991).}

\bibitem{arimondo_J0}
{H.~Lignier, C.~Sias, D.~Ciampini, Y.~Singh, A.~Zenesini,
O.~Morsch, and E.~Arimondo, Phys. Rev. Lett. {\bf 99}, 220403 (2007).}

\bibitem{arimondo_Jn}
{C.~Sias, H.~Lignier, Y.P.~Singh, A.~Zenesini, D.~Ciampini,
O.~Morsch, and E.~Arimondo, Phys. Rev. Lett. {\bf 100}, 040404 (2008).}

\bibitem{cec_morsch}
{C.E.~Creffield, F.~Sols, D.~Ciampini, O.~Morsch, and E.~Arimondo.
Phys. Rev. A {\bf 82}, 035601 (2010).}

\bibitem{cec_J0}
{C.E.~Creffield and F.~Sols, Phys. Rev. Lett. {\bf 100}, 250402 (2008).}

\bibitem{cec_phase}
{C.E.~Creffield and F.~Sols, Phys. Rev. A {\bf 84}, 023630 (2011).}

\bibitem{longhi}
{S.~Longhi, Opt. Lett. {\bf 38}, 3570 (2013).}

\bibitem{eckardt1}
{J.~Struck, C.~\"Olschl\"ager, M.~Weinberg, P.~Hauke, J.~Simonet, A.~Eckardt,
M.~Lewenstein, K.~Sengstock, and P.~Windpassinger,
Phys. Rev. Lett. {\bf 108}, 225304 (2012).}

\bibitem{eckardt2}
{P.~Hauke, O.~Tieleman, A.~Celi, C.~\"Olschl\"ager, J.~Simonet, J.~Struck,
M.~Weinberg, P.~Windpassinger, K.~Sengstock, M.~Lewenstein, and A.~Eckardt,
Phys. Rev. Lett. {\bf 109}, 145301 (2012).}

\bibitem{kolovsky}
{A.R.~Kolovsky, Europhys. Lett. {\bf 93}, 20003 (2011).}

\bibitem{comment}
{C.E.~Creffield and F.~Sols, Europhys. Lett. {\bf 101}, 40001 (2013).}

\bibitem{ketterle}
{H.~Miyake, G.A.~Siviloglou, C.J.~Kennedy, W.C.~Burton, and W.~Ketterle,
Phys. Rev. Lett. {\bf 111}, 185302 (2013).}

\bibitem{bloch}
{M.~Aidelsburger, M.~Atala, M.~Lohse, J.T.~Barreiro, B.~Paredes, and I.~Bloch,
Phys. Rev. Lett. {\bf 111}, 185301 (2013).}

\bibitem{troyer}
{L.~Wang and M.~Troyer, Phys. Rev. A {\bf 89}, 011603(R) (2014).}

\bibitem{holthaus_Jn}
{A.~Eckardt, T.~Jinasundera, C.~Weiss, and M.~Holthaus,
Phys. Rev. Lett. {\bf 95}, 200401 (2005).}

\bibitem{innsbruck}
{E.~Haller, R.~Hart, M.J.~Mark, J.G.~Danzl, L.~Reichs\"ollner, and
H.-C.~N\"agerl, Phys. Rev. Lett. {\bf 104}, 200403 (2010).}

\bibitem{tania_phase}
{K.~Kudo and T.S.~Monteiro, Phys. Rev. A {\bf 83}, 053627 (2011).}

\bibitem{footnote_phase}
{This cancellation of phases is a general feature of lattices
with parallel-sided plaquettes, such as square and
hexagonal lattices. It does not, however, apply to triangular
lattices, which is why a non-zero (albeit staggered) flux could
be observed in Ref. \onlinecite{eckardt2}.}

\bibitem{soren}
{A.S.~S{\o}rensen, E.~Demler, and M.D.~Lukin, Phys. Rev. Lett. {\bf 94}, 
086803 (2005).}
  
\bibitem{footnote}
{This form of flipping the sign of the acceleration was
used in the resonant driving experiments of Ref. \onlinecite{arimondo_Jn}.}

\bibitem{suzuki}
{M. Suzuki, Phys. Lett. A {\bf 146}, 287 (1990).}

\bibitem{kol_new}
{A.R.~Kolovsky, F.~Grusdt, and M. Fleischhauer, Phys. Rev. A {\bf 89},
033607 (2014).}

\bibitem{slosh}
{S.~Burger, F.S.~Cataliotti, C.~Fort, F.~Minardi, M.~Inguscio, M.L.~Chiofalo,
and M.P.~Tosi, Phys. Rev. Lett. {\bf 86}, 4447 (2001).}

\bibitem{hanggi}
{M.~Grifoni and P.~H\"anggi, Phys. Rep. {\bf 304}, 229 (1998).}

\bibitem{belen}
{D.~H\"ugel and B.~Paredes, arXiv:1306.1190}

\bibitem{demler}
{T.~Kitagawa, E.~Berg, M.~Rudner, and E.~Demler,
Phys. Rev. B {\bf 82}, 235114 (2010).}

\end{thebibliography}
\end{document}